\documentclass[reprint,amsmath,amssymb,
 aps, superscriptaddress,nofootinbib,longbibliography]{revtex4-1}
\usepackage{amsmath, braket, graphicx, amssymb}
\usepackage{bm,bbm}
\usepackage{hhline}
\usepackage{dcolumn}
\usepackage[english]{babel}
\usepackage{physics}
\usepackage{bbold}
\usepackage{mathtools}
\usepackage{xcolor}
\usepackage{soul}

\newcommand{\tcb}{}

\usepackage{natbib}
\setcitestyle{round,sort}

\begin{document}
\date{\today}
\title{Qubit Readout Error Mitigation with Bit-flip Averaging}
\author{Alistair W. R. Smith}
\affiliation{QOLS, Blackett Laboratory, Imperial College London SW7 2AZ, United Kingdom}
\email{Corresponding author: alistair.smith18@imperial.ac.uk}
\author{Kiran E. Khosla}
\affiliation{QOLS, Blackett Laboratory, Imperial College London SW7 2AZ, United Kingdom}
\author{Chris N. Self}
\affiliation{QOLS, Blackett Laboratory, Imperial College London SW7 2AZ, United Kingdom}
\author{M. S. Kim}
\affiliation{QOLS, Blackett Laboratory, Imperial College London SW7 2AZ, United Kingdom}
\begin{abstract}
Quantum computers are becoming increasingly accessible, and may soon outperform classical computers for useful tasks. However, qubit readout errors remain a significant hurdle to running quantum algorithms on current devices. We present a scheme to more efficiently mitigate these errors on quantum hardware and numerically show that our method consistently gives advantage over previous mitigation schemes. Our scheme removes biases in the readout errors allowing a general error model to be built with far fewer calibration measurements. Specifically, for reading out $n$-qubits we show a factor of $2^n$ reduction in the number of calibration measurements without sacrificing the ability to compensate for correlated errors. Our approach can be combined with, and simplify, other mitigation methods allowing tractable mitigation even for large numbers of qubits.

\end{abstract}
\maketitle
\section{Introduction}\label{sec:intro}
\tcb{Noisy-intermediate scale quantum (NISQ) computers \cite{Preskill2018} are running increasingly complicated algorithms on small to intermediate numbers of qubits \citep{Kandala_2017, Nam2020Apr, 2020Aug, Smith2019Nov, Havlicek2019, Johri2021, Arute2019,Hu2020,Vovrosh2021}.} However, their usefulness continues to be limited by noise, leading to unreliable outputs. \tcb{Error mitigation schemes \citep{Endo2018Jul,Endo2021Feb} compensate for errors through a combination of calibration measurements and post-processing and have been applied to the benchmarking of NISQ hardware \citep{mit1,mit2,mit5,bravyi_2020}, quantum chemistry and solid state physics problems \cite{McCaskey2019,Kandala2019,Self2021}, dynamical quantum simulations \citep{mit2,Vovrosh2021}, and demonstrations of  quantum  supremacy \citep{2020Aug}.} They have been proposed to bridge the gap between current devices and future fault-tolerant error correction \cite{campbell_2017}, which actively corrects errors in the quantum state. Error mitigation has already proven to be an important tool to reach new benchmarks on existing hardware \cite{Kandala_2017, Zalcman2020}.    

Qubit readout is a significant source of error in quantum computing experiments. This is particularly true for the popular superconducting qubit architectures, \tcb{which typically have per-qubit readout error probabilities of a few percent}  \tcb{\cite{Arute2019} (detailed information about readout error probabilities on current devices can be found through IBM's Qiskit platform \cite{qiskit})}.
In practice the measurement errors on transmon-based devices are additionally complicated by effects such as bias towards certain states and cross-talk induced correlations \cite{meas_asym,cross_talk1,cross_talk2}.
Furthermore, quantum experiments often involve measuring many qubits at a time \cite{McArdle2020Mar,Cleve1998Jan}, compounding the impact of readout errors. Together, these effects make readout errors a significant hurdle to scaling up NISQ computation.

Readout error mitigation schemes combine an error model with calibration measurements. The calibrated model is then used to infer the ``error-free" result of an experiment~\cite{classical_model,bit_flip_expect,cross_talk1,state_bias,Kwon2020Jul,Kim2020Oct,Endo2018Jul,Zlokapa2020May,Czarnik2020May,Harrigan2021,Czarnik2021,rebalance,Berg2021,Wang2021,peters2021,Maciejewski2021}. The quality of the mitigation strongly depends on the choice of error model, however there is a trade-off between model complexity and calibration cost. Simple models, for example those assuming qubit-wise-independent errors \citep{Kandala_2017,Arute2019}, require fewer calibration measurements but may not capture the true error process. In contrast, using fewer assumptions leads to a more general error model but at the cost of potentially requiring a prohibitive number of calibration measurements \cite{qiskit}. Here we present a scheme that addresses both these problems, giving a lossless reduction in error model complexity and introducing a single, model agnostic, calibration step. This allows the most suitable model to be chosen \emph{a posteriori}.

Here we introduce bit-flip averaging (BFA), a scheme that uses random bit-flips to simplify the effective error process. We analytically show that averaging over these random bit-flips, allows one to more efficiently parameterise, and estimate readout errors. The error process under BFA admits convenient mathematical symmetries that greatly simplifies the inference of ``error-free" experimental results. We compare our approach to full mitigation and tensor product noise (TPN) models, and show that BFA outperforms both. The bit-flips introduced by our method can be uniquely inverted, allowing for mid-circuit measurement and feed-forward algorithms \citep{qiskit,Cong2019Dec} experimental overhead, requiring only a layer of single qubit gates and classical post-processing. 

Imperfect multi-qubit measurements can be effectively modeled as a classical process \cite{bravyi_2020,geller_2020,hamilton_2020,classical_model,classical_model2}.
This can be understood as a probabilistic corruption of the error-free result.
Assuming that the measurements (in the computational basis) will be performed across a constant number of qubits, this model is expressed in terms of a response matrix $\mathbf{M}$ such that $M_{\sigma \sigma'}=p_{(\sigma|\sigma')}$, 
gives the probability of reading out $\sigma$ given that the error-free outcome should have been $\sigma'$. The observed outcome probabilities $\boldsymbol{p}_{\mathrm{obs}}$ are given by the action of the response matrix on the error-free probabilities $\boldsymbol{p}_{\mathrm{true}}$,
\begin{equation}\label{rmat}
\boldsymbol{p}_{\mathrm{obs}}=\mathbf{M}\boldsymbol{p}_{\mathrm{true}}.
\end{equation}
In general the matrix $\mathbf{M}$ is not symmetric as readout on many devices is biased towards some states \cite{meas_asym}. 
Our protocol uses random bit-flips to symmetrise the response matrix, averaging out the biases. This drastically reduces the number of parameters required to define this matrix; this reduction is $\mathcal{O}(2^{2n})\to\mathcal{O}(2^n)$ for $n$ read-out qubits. It also simplifies the matrix inversion task required to find $\boldsymbol{p}_{\mathrm{true}}$.\par

The calibration step involves estimating $\mathbf{M}$ by preparing and measuring each of the computational basis states. The $k$th column of $\mathbf{M}$ is the vector of measurement outcome probabilities given an input computational basis state $\ket{k}$. 
This requires enough calibration shots to sufficiently determine $2^n$ (potentially) unique probabilities for each of the $2^n$ different $\ket{k}$, which is especially problematic if time-drifting errors necessitate frequent re-calibrations. At worst, calibration costs scale as $\mathcal{O}(2^{2n})$, however, in practice many of the error probabilities will be negligibly small (i.e. those for simultaneous errors on many qubits) and can be safely approximated as zero. We show in section~\ref{subsec:scaling} that the number of calibration measurements needed to estimate a single distribution (column of $\mathbf{M}$) typically scales at a rate $\ll\mathcal{O}(2^n)$ (although still exponentially in $n$). Nevertheless, even if each distribution can be described with a dramatically reduced set of probabilities, there are still exponentially many distributions (input states) to estimate.

Once $\mathbf{M}$ is estimated, readout errors are typically mitigated by either inverting $\mathbf{M}$ or by solving a constrained linear optimisation problem (minimizing $(\boldsymbol{p}_{\mathrm{obs}}-\mathbf{M}\boldsymbol{p})^2$ over $\boldsymbol{p}$, subject to physical probabilities). We note that both problems quickly become intractable with increasing numbers of qubits.\par

In practice, there will be some underlying structure to the readout error distributions. Several proposals have taken advantage of this by making assumptions about the error process \citep{bravyi_2020,classical_model,state_bias,bit_flip_expect}. A common and effective choice of simplified model assumes that the readout errors for each qubit are independent, yielding the so-called tensor product noise (TPN) model \cite{bravyi_2020,Geller2020Jun,Kandala_2017}. This simplification allows the response matrix to be given in terms of $2n$ single qubit error probabilities $\{p^{i}_{(1|0)}\}$ and $\{p^{i}_{(0|1)}\}$. For TPN, the response matrix $\mathbf{M}_{\text{TPN}}$ is the tensor product of single-qubit response matrices,\par
\begin{equation}\label{tpn_eq}
\mathbf{M}_{\text{TPN}}=\bigotimes_{i}
\begin{pmatrix}
p^i_{(0|0)} & p^i_{(0|1)}\\
p^i_{(1|0)} & p^i_{(1|1)}
\end{pmatrix},
\end{equation}
where $p^{i}_{(s|s')}$ is the probability that the $i$th qubit reads out $s$ given the error-free readout should have been $s'$ (and $p^{i}_{(0|0)} = 1 - p^{i}_{(1|0)}$ etc.). The TPN model can be calibrated more efficiently than the full scheme as $\{p^i_{(1|0)}\}$ and  $\{ p^i_{(0|1)} \}$ can be found by sampling only the input states $\ket{0\dots 0}$ and $\ket{1\dots 1}$ respectively. The inverse of $\mathbf{M}_{\text{TPN}}$ is now tractable, and is simply the tensor product of inverse single-qubit response matrices.  

On real devices multi-qubit readout errors can be correlated \cite{bravyi_2020,meas_asym} (through cross-talk effects) limiting the accuracy of many simplified models. Alternative approaches have been proposed to deal with correlated errors in a scalable way, e.g. using continuous Markov processes \cite{bravyi_2020}, or via cumulant expansion \citep{hamilton_2020}. These methods extend the TPN approximation by characterising the readout errors in terms of single qubit and pair-wise (between physically/frequency close qubits) correlated error terms. Although we do not consider these models directly in this paper, our BFA proposal naturally extends to these correlation-extended models. Furthermore, as our scheme eliminates the bias in the readout errors towards certain states it allows for these models to be expressed in terms of fewer parameters. This simplification allows these models to be calibrated with fewer measurements and thereby mitigate readout errors more efficiently.
\begin{figure*}
\includegraphics[scale=0.33]{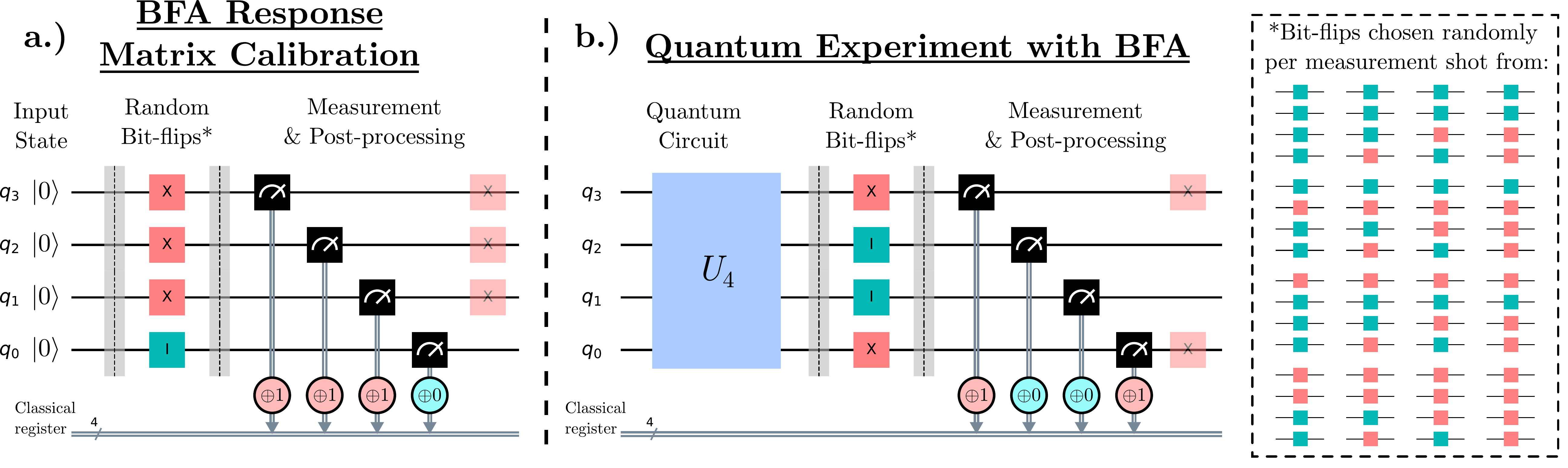}
\caption{\textbf{Example of bit-flip averaging for 4 qubits.} \textbf{a.)} Determining the response matrix requires applying $ \hat{X}$-gates to randomly chosen qubits (red squares) while leaving others unflipped (cyan squares). The bit-flipping is inverted in the classical readout result (red and cyan circles) leaving the resulting measurement invariant under the full process. Averaging over many random choices of bit-flipped qubits means that the effective readout error process is uniquely determined by considering only the logical zero state. \textbf{b.)} In order to use the BFA response matrix to mitigate errors in an experiment, it is crucial the experiment sees the same effective measurement process, this is simply achieved by applying BFA to the experimental measurement process. If required, the $\hat{X}$ gates can be inverted (transparent squares), leaving the quantum state invariant under BFA.}
\label{fig:bfa_diag}
\end{figure*}

\section{Results}
\subsection{Averaging-out Readout Errors with Bit-flips}\label{subsec:flip}
Our BFA method yields a greatly simplified and more easily measurable response matrix without sacrificing the ability to capture correlated readout errors. By applying random pre- and post-measurement bit-flips, we completely symmetrise the response matrix and remove readout biases. The process is qualitatively similar to the randomised benchmarking techniques \cite{Knill2008} that are often used to efficiently quantify gate errors. Methods that tackle state dependent bias have been proposed, e.g. the ``Static Invert-and-Measure" scheme \cite{state_bias}, however the scheme introduced here allows for more active and efficient response-matrix-based mitigation to be used. Our scheme also provides computational advantages in applying the mitigation; we give an analytic formula for the inverse of the simplified response matrix under BFA that can be calculated using only vector-matrix multiplication.\par
Following the standard response matrix approach, we assume that every measurement will be performed across a fixed number of qubits. Mitigation happens in two stages; a calibration stage where the response matrix is measured, and an experimental stage in which readout errors are mitigated using this response matrix. In each shot of the experiment we bit-flip random qubits before measuring them and then invert the bit-flip in the (classical) readout for the corresponding qubits (Figure~\ref{fig:bfa_diag}). We repeat and average over this process, randomly selecting different qubtis to bit-flip each shot. As we will show, this simplifies the measured effective response matrix (measurement of this is shown in Figure~\ref{fig:bfa_diag}a). The bit-flipped qubits are chosen uniformly at random per shot and the bit-flips are implemented with an $\hat{X}$ gate. Here and in the following, we assume the gate errors introduced by $\hat{X}$ negligible. By continuing the random bit-flips and classical correction when performing a quantum experiment the readout errors continue to be simplified, allowing for easier mitigation (Figure~\ref{fig:bfa_diag}b). In the absence of any readout errors, BFA has no effect on the readout results. Finally, if necessary, we can include another set of $\hat{X}$ gates to the post-measurement flipped qubits ensuring a consistent \textit{a posteriori} state.\par

Without loss of generality, we describe the measurement operation, and corresponding response matrix using Kraus operators. For $n$ qubits we do this in terms of a set of $2^n$ measurement Kraus operators $\{A_{\sigma}\}$ where
\begin{equation}\label{kraus1}
A_{\sigma}=\sum_{\sigma'}\sqrt{p_{(\sigma|\sigma')}}\ketbra{\sigma'}{\sigma'},
\end{equation}
which was chosen in such a way to recover Eq.~\ref{rmat}.
This operator $A_{\sigma}$ corresponds to an uncertain measurement (with $\{\ket{\sigma'}\}$ being computational basis vectors), yielding a classical readout bit-string $\sigma$. We note that $A_{\sigma}$ corresponds to a quantum noise limited measurement operator. An additional incoherent classical assignment error can be included, however this also gives a response matrix of the form in Eq.~\ref{rmat}. Our BFA scheme yields the same result for any combination of these two processes.

Eq.~\ref{rmat} is recovered by considering the probability $\boldsymbol{p}_{obs}(\sigma)=\Tr{A^\dag_{\sigma} A_{\sigma}\rho}$ that a measurement of a quantum state $\rho$ yields a readout $\sigma$; 
\begin{equation}
    \begin{split}
        \boldsymbol{p}_{\mathrm{obs}}(\sigma) &=\sum_{\sigma'}p_{(\sigma|\sigma')}\boldsymbol{p}_{\mathrm{true}}(\sigma') 
    \end{split}
\end{equation}
where $\boldsymbol{p}_{\mathrm{true}}(\sigma')=\bra{\sigma'}\rho\ket{\sigma'}$ is the probability of an error-free measurement of $\rho$ to yield the state $\ket{\sigma'}$ 
We identify this sum as the matrix equation $\boldsymbol{p}_{\mathrm{obs}}=\mathbf{M}\boldsymbol{p}_{\mathrm{true}}$.\par
BFA can be taken into account by applying the relevant bit-flipping operations and adjusted readout results directly to the measurement operators. 
We quantify a bit-flip in terms of a binary string $s$ such that the unitary operator applied to the qubits is $\hat{X}^{(s)}=\bigotimes_i\hat{X}^{s_i}$, a tensor product of Pauli $\hat{X}$ operators and identity operators where $s_i$ is the $i$th bit of $s$, e.g. $\hat{X}^{(01)}=\mathbb{1}\otimes\hat{X}$.
For an $n$-qubit measurement we choose a random bit-string $s$ with probability $1/2^n$, and given a bit-string readout $\sigma$ the corresponding measurement operator is:
\begin{equation}
\begin{split}\label{flipop}
A_{s, \sigma}&=
\frac{1}{\sqrt{2^n}}\sum_{\sigma'}\sqrt{p_{(\sigma\oplus s|\sigma')}} \ketbra{\sigma'\oplus s}{\sigma'\oplus s}.
\end{split}
\end{equation}
where $\ket{\sigma'\oplus s} = \hat{X}^{(s)} \ket{\sigma'}$ and $\sigma'\oplus s$ is the bit-wise addition of $s$ and $\sigma'$. 
We now consider how averaging over $s$ changes the readout error process. From here, averaging over $s$ is denoted by a tilde. The ($s$-averaged) probability of observing $\sigma$ is,
\begin{equation}\label{flipmeas}
\begin{split}
        \boldsymbol{\tilde{p}}_{\mathrm{obs}}(\sigma) 
        &=
        \frac{1}{2^n}\sum_{s,\sigma'}p_{(\sigma\oplus s|\sigma'\oplus s)}\boldsymbol{p}_{\mathrm{true}}(\sigma'),
\end{split}
\end{equation}
allowing us to identify a new response matrix $\widetilde{\mathbf{M}}$ (such that $\boldsymbol{\tilde{p}}_{\mathrm{obs}}=\widetilde{\mathbf{M}}\boldsymbol{p}_{\mathrm{true}}$) that describes the readout errors under BFA. The elements of this new matrix are simply the $s$-averaged conditional probabilities $\tilde{p}_{(\sigma|\sigma')}$ under BFA, and are given by:
\begin{equation}\label{eq:ptilde}
   \widetilde{M}_{\sigma\sigma'}=\frac{1}{2^n}\sum_{s} p_{(\sigma\oplus s|\sigma'\oplus s)}.
\end{equation}
From this equation, we see that $\tilde{p}_{(\sigma\oplus s|\sigma'\oplus s)}=\tilde{p}_{(\sigma|\sigma')}$ and so we have arrived at a \textit{far simpler, symmetrised error model} with $2^n\!-\!1$ parameters, instead of the $2^n(2^n\!-\!1)$ in $\mathbf{M}$.\par
Thanks to this symmetry, we can express the whole response matrix under this bit-flipping protocol $\widetilde{\mathbf{M}}$ in terms of just the parameters in its first column. 
The conditional index in the response matrix can now be dropped as 
\begin{equation}
\widetilde{M}_{\sigma\sigma'}=\tilde{p}_{(\sigma\oplus \sigma'|0)} \equiv \tilde{p}_{(\sigma\oplus \sigma')}.
\label{eq:M_BFA}
\end{equation}
Physically this is because any computational basis state is equally likely to be bit-flipped into any other basis state. As it is the bit-flipped state that is measured by the physical apparatus, the effective error probability is averaged across all inputs, removing any bias towards certain states. This removal of bias gives a huge practical advantage of BFA over normal response matrix error mitigation; calibrating the entire symmetrised error matrix only requires measurement of the probabilities in its first column which is done with just \textit{one} input state $\ket{0\dots 0}$. As no assumptions are made about whether the errors are correlated, such correlations can be effectively dealt with using this scheme. In particular, one can infer correlations by comparing the relative probabilities of different errors in the calibration data.\par

\tcb{For practical reasons, readout errors are often biased towards certain states. With superconducting qubits, the readout of qubits in the state $\ket{0}$ is typically more reliable than for the state $\ket{1}$. Adaptive mitigation schemes \cite{state_bias,rebalance} have been proposed to more effectively compensate for readout errors by exploiting this bias. These involve applying $\hat{X}$ gates to certain qubits before measurement to maximise the proportion that are measured in the state $\ket{0}$, reducing the probability of a readout errors and allowing the error-corrupted measurement distributions to be estimated more easily. However, these schemes only yield improvements for certain states and require the circuits for suitable states to be run twice (without and with adaptive $\hat{X}$ gates) while also requiring the full response matrix to be found. By symmetrizing the readout errors BFA increases the probability of some states (those with a high proportion of $\ket{0}$'s) being read out incorrectly but this is balanced by effective error probability for other states (with a high proportion of $\ket{1}$'s) being reduced. The balancing of readout errors coupled with the factor of $2^n$ reduction in the cost of estimating the response matrix and the lack of circuit-specific calibration measurements allows BFA to compete with these adaptive schemes while being applicable to a broader set of states.}

The probabilities $\tilde{p}_{(S)}$ have a convenient physical interpretation as the average probability that an error with syndrome $S$ (the bit-string identifying which qubits are read out incorrectly) occurs, e.g. $\tilde{p}_{(1011)}$ is the probability that readout errors occur simultaneously on the $0^{th}$, $1^{st}$ and $3^{rd}$ qubits. The error matrix $\widetilde{\mathbf{M}}$ is symmetric about both its diagonal and anti-diagonal, allowing it to be decomposed into a compact form,
\begin{equation}\label{sparse}
\widetilde{\mathbf{M}}=\sum_s \tilde{p}_{(s)}X^{(s)},
\end{equation}
where we have used the same notation for the matrix $X^{(s)}$ as for the operator $\hat{X}^{(s)}$. This is due to Eq.~\ref{eq:ptilde} being invariant under $\sigma,\sigma'\to \sigma\oplus s,\sigma'\oplus s$ (i.e. invariant under $\widetilde{\mathbf{M}}\to X^{(s)}\widetilde{\mathbf{M}}X^{(s)}$).
To take advantage of this sparse representation one must continue to perform the bit-flipping and classical correction during experiments. This requirement adds negligible overhead as single qubit bit-flips can typically be performed with very high fidelity, or are completely free if combined with an existing gate. The corresponding bit-flip of the measurement output requires only classical Boolean logic.\par
The decomposition in Eq.~\ref{sparse} gives us an advantage in both mitigation strategies (response matrix inversion and constrained least squares minimization). For the least squares method, Eq.~\ref{sparse} tells us with what probability $\tilde{p}_{(S)}$ we expect a given readout string to be corrupted by the binary addition of $S$. If many of these probabilities are zero (or negligibly small and so can be set to zero) then this would allow us to use a sparse matrix representation of $\widetilde{\mathbf{M}}$, allowing the optimisation problem to be solved more easily.\par
The matrix inverse mitigation strategy requires $\widetilde{\mathbf{M}}^{-1}$ to be found. As $\widetilde{\mathbf{M}}$ contains only tensor products of the identity and Pauli $X$ matrix all these terms can be simultaneously diagonalised by the application of the Hadamard matrix $\text{H}$. As we show in section~\ref{deriv}, this gives the vector of eigenvalues
\begin{equation}\label{eigs}
\lambda=\sqrt{2^n}\text{H}^{\otimes n}\tilde{p},
\end{equation}
where $\tilde{p}$ is a vector of the probabilities $(\tilde{p})_s\equiv \tilde{p}_{(s)}$ (corresponding to the first column of $\widetilde{\mathbf{M}}$). Like $\widetilde{\mathbf{M}}$, the inverse must be symmetric about both its diagonal and anti-diagonal meaning that it can also be decomposed onto Pauli $X$ matrices, i.e. the form given in Eq.~\ref{sparse}. As shown in section~\ref{deriv}, the inverse BFA-simplified response matrix is given in terms of the vector of reciprocal eigenvalues $\lambda^{-1}\equiv (1/\lambda_0, \dots, 1/\lambda_{n-1})$ by
\begin{equation}\label{sparse_inv}
\widetilde{\mathbf{M}}^{-1}=\sum_s \tilde{q}_{(s)} X^{(s)}, \quad \tilde{q}=\frac{1}{\sqrt{2^n}}\text{H}^{\otimes n}\lambda^{-1}.
\end{equation}
This shows another clear advantage to bit-flipping over the full mitigation approach. Like with $\widetilde{\mathbf{M}}$, we only need to find the elements of the inverse's first column and this can be done with simple matrix multiplication (as opposed to a computationally costly general matrix inverse).\par
As the primary function of BFA is to average out bias in the readout errors towards certain measurement outcomes we can also use it to further simplify other simplified error models. As an example of how BFA can simplify other approximate measurement error mitigation protocols we can consider how the TPN model transforms under bit-flipping. Under bit-flipping, the biases of the qubit-wise readout errors are averaged out meaning that the BFA-symmetrised TPN matrix $\widetilde{\mathbf{M}}_{\text{TPN}}$ is given by
\begin{equation}
\widetilde{\mathbf{M}}_{\text{TPN}}=\bigotimes_{i}
\begin{pmatrix}
1-\tilde{p}_i & \tilde{p}_i\\
\tilde{p}_i & 1-\tilde{p}_i
\end{pmatrix},\ \tilde{p}_{i} = \frac{p^i_{(1|0)} + p^i_{(0|1)}}{2}.
\end{equation}
Combining the TPN model with BFA provides two main advantages; the first being that the number of parameters to estimate for the combined model on $n$ qubits is $n$ instead of $2n$. The second comes in estimating $\{\tilde{p}_i\}$; these probabilities can be measured by preparing the state $\ket{0\dots 0}$. This is the same experimental procedure as is required for calibration of the full BFA matrix $\widetilde{\mathbf{M}}$ and so a single set of calibration results can be used for both models. In this example, a TPN+BFA model could be calibrated first and its predictions for the different error probabilities checked against the calibration data. If this proves unsuitable then a larger more general model could be employed without requiring any further calibration measurements. 

The information contained in the BFA calibration measurements of $\ket{0\dots0}$ fully describes the response matrix and so it can be used to calibrate any \emph{any} response matrix based approach. This means that one is not forced to make any assumptions about the model (e.g. independent errors, pairwise correlations, or a full model) before calibration. 

This flexibility potentially allows for readout error mitigation to be performed even for large numbers of qubits, provided any correlations in the readout errors have some degree of locality. If the qubits can be grouped into disjoint sets such that there are no inter-group correlations (for example if correlations only occur between qubits coupled to the same readout cavity) then an expanded TPN-like model could be used in which each group has its own full response matrix. The response matrix for a measurement of all the qubits would then be given by a tensor product of those for each group. BFA would allow this model to be calibrated using only a single measured input state at a cost scaling at worst as $\mathcal{O}(2^k)$ where $k<n$ is the number of qubits in the largest grouping. While the groupings could initially be chosen based on some knowledge of the device (e.g. by readout cavity, operating frequency, or some spatial consideration), BFA would allow this grouping to be changed retroactively to match the calibration data. \tcb{In the supplementary text we show an example calibration process for a sparse response matrix that exhibits correlations between only a restricted number of qubits. We demonstrate how an appropriate choice of model can make both the calibration and subsequent mitigation more accurate.} 

\subsection{Sample complexity and scaling of full BFA calibration}\label{subsec:scaling}
To use BFA most effectively it should be combined with an error model that best balances the trade-off between model expressibility and calibration cost. However, it is helpful to estimate the worst-case cost to calibrate a BFA model using the full symmetrised response matrix as in Eq.~\ref{sparse}, under some physically motivated assumptions. On real devices we expect that even when correlations are taken into account, the probabilities of errors occurring on many qubits simultaneously is negligible and can be neglected. This effectively reduces the number of parameters that must be estimated to find $\tilde{p}$ and thereby $\widetilde{\mathbf{M}}$. The number of probabilities that give non-negligible contributions to $\tilde{p}$ will provide an indication of the number of calibration measurements required to estimate $\widetilde{\mathbf{M}}$, and equivalently the cost to estimate a single column of $\mathbf{M}$.

To give an idea of how many parameters must be retained in $\tilde{p}$ we consider a TPN model for an $n$-qubit readout with constant single-qubit readout error probabilities $p_e$ for all qubits. While this neglects any correlations in the errors, we expect that these would act as a modest correction to the TPN model and so would not greatly impact the calibration cost. For a conservative scaling estimate $p_e$ could be the largest measured single-qubit readout error probability for the device in question. We then calculate the number $N$ of error probabilities that must be retained so that their cumulative probability reaches above a threshold $\sum_{i=0}^{N-1}\tilde{p}_{(i)}^{\downarrow}>1-\epsilon$ (where $\tilde{p}^{\downarrow}$ is $\tilde{p}$ sorted in descending order). 

Under these assumptions the number of qubits $Q$ that experience a readout error is binomially distributed $Q \sim \text{B}(n,p_e)$. From this we can calculate the highest weight of error $k$ (i.e. largest number qubits that are simultaneously read out incorrectly) that must be retained such that the cumulative probability up to $k$ is greater than $1-\epsilon$. This is given by $k=S^{-1}_{B}(\epsilon;n,p_e)$, where $S_{B}(k;n,p_e)=1-Pr(Q\leq k)$. Thus, $N$ is the number of possible readout errors with weight less than or equal to $k$;
\begin{equation}
    N=\sum_{i=0}^{k}\binom{n}{i}.
\end{equation}
While this sum does not have a closed form solution, for $k/n\leq1/2$ (expected for large $n$ and $p_e\ll 1$) it is bounded \cite{Ash_1965} by 
\begin{equation}\label{bounds}
    \frac{1}{\sqrt{8k(1-k/n)}}2^{nH(k/n))}\leq N\leq 2^{nH(k/n)},
\end{equation}
where $H(p)=-p\log_2(p)-(1-p)\log_2(1-p)$ is the binary entropy function. This means the number of measurements needed to calibrate a completely general $\widetilde{\mathbf{M}}$ is expected to scale at worst as $\mathcal{O}(2^{nH(k/n)})$. Although the number of outcomes that significantly contribute to the total probability is typically much less than the $2^n$ worst case, the required number of terms still scales exponentially with $n$. 

For large $n$ we can examine the scaling of $N$ by approximating $Q$ with a normal distribution $Q\sim\mathcal{N}(np_e,np_e(1-p_e)$ giving
\begin{equation}
k\approx np_e + \sqrt{np_e(1-p_e)}S^{-1}_{\mathcal{N}}(\epsilon)+0.5,
\end{equation}
where $S_{\mathcal{N}}(k)=1-\text{Erf}(k)$, and we have applied a continuity correction of $0.5$. Using this approximation we see that in the limit of large $n$ the lower bound tends to $2^{nH(p_e)}/\sqrt{8np_e(1-p_e)}$, and so we require \emph{at least} $\mathcal{O}(2^{(nH(p_e))})$ samples to calibrate $\widetilde{\textbf{M}}$. As the upper bound on $N$ tends to $2^{nH(p_e)}$ we can also identify a very rough rule-of-thumb for the errors rates under which full mitigation is tractable, e.g. $nH(p_e)<10$.

\tcb{We can further quantify the sample complexity of estimating a typical $\widetilde{\mathbf{M}}$ by applying bounds for the estimation of arbitrary discrete distributions. A useful bound for this is given in \cite{Berend2012} and states that for a $N$-outcome discrete distribution with true probabilities $\boldsymbol{p}$, the expected total variation distance $\delta(\boldsymbol{p},\hat{\boldsymbol{p}}^{(m)})=\sum_i|p_i-\hat{p}^{(m)}_i|$ between $\boldsymbol{p}$ and the empirical distribution after $m$ samples $\hat{\boldsymbol{p}}^{(m)}$ is bounded by:
\begin{equation}
    \mathbb{E}[\delta(\boldsymbol{p},\hat{\boldsymbol{p}}^{(m)})]\leq\sqrt{\frac{N}{m}}.
\end{equation}
This shows that we can estimate $\boldsymbol{p}$ to an expected accuracy $\epsilon$ using $m=N/\epsilon^2$ shots. From \cite{Berend2012} we also have a concentration bound for this inequality. The probability $P[\delta(\boldsymbol{p},\hat{\boldsymbol{p}}^{(m)})>\epsilon]$ that we observe a distribution after $m$ shots that differs from the true distribution with $\delta>\epsilon$ is bounded for $\epsilon\geq\sqrt{N/m}$ by:
\begin{equation}
    P[\delta(\boldsymbol{p},\hat{\boldsymbol{p}}^{(m)})>\epsilon]\leq\exp\left(-\frac{m}{2}\left(\epsilon-\sqrt{\frac{N}{m}}\right)^2\right).
\end{equation}
This implies that to be sure that we have estimated $\boldsymbol{p}$ to within an accuracy $\epsilon$ with a failure probability less than $\gamma$ we need at least $m>(\sqrt{N}+\sqrt{2\log\frac{1}{\gamma}})^2/\epsilon^2$ shots.}

\tcb{Therefore, estimating an arbitrary $n$-qubit response matrix using BFA to an accuracy $\epsilon$ with failure probability less than $\gamma$ requires at least $\tilde{m}(n,\epsilon,\gamma)\equiv(2^{(n/2)}+\sqrt{2\log\frac{1}{\gamma}})^2/\epsilon^2$ or $\mathcal{O}(2^n/\epsilon^2)$ shots. With full mitigation all $2^n$ columns must be estimated independently meaning a worst-case sample cost of $2^n\times\tilde{m}(n,\epsilon,\gamma)$ and an $\mathcal{O}(2^{2n}/\epsilon^2)$ sample complexity.} 

\tcb{These worst-case sample complexities correspond to a completely general $\mathbf{M}$ in which all $2^n$ outcomes in each column are significant. From Eq.~\ref{bounds}, the typical number of significant probabilities $N$ given a representative single-qubit readout error probability $p_e$ is bounded by $N\leq2^{nH(k/n)}$ (where $k$ is defined as before). This yields to give a more typical estimate for the sampling cost under BFA of $\tilde{m}(nH(k/n),\epsilon,\gamma)$ shots and, for large $n$, a sampling complexity of $\mathcal{O}(2^{nH(p_e)}/\epsilon^2)$ shots.}

\tcb{The exact sampling cost depends on the values of the various error probabilities. In the above analysis we assume that the sampling cost under BFA can be reasonably well-approximated by the cost of naively sampling a symmetric TPN response matrix with unbiased and equal per-qubit readout errors. While this is a reasonable approximation for BFA it ignores the readout error biases encountered in full mitigation schemes. With the typical biases observed for superconducting qubits, the columns (output distributions) corresponding to input states with large numbers of qubits in the state $\ket{0}$ are more sharply peaked than for states with many qubits in $\ket{1}$ (as the former will have lower readout error probabilities). Sharper distributions (e.g. $p_{(\sigma|0\dots0)}$) can be estimated with fewer samples than flatter ones (e.g. $p_{(\sigma|1\dots1)}$) so the required number of samples to reach a given accuracy will vary across the columns of $\mathbf{M}$. Due to the averaging that occurs, the sample complexity of estimating the single distribution $\tilde{p}_{(\sigma)}$ required for BFA will usually lie somewhere between the best-case cost of estimating $p_{(\sigma|0\dots0)}$ and worst-case of $p_{(\sigma|1\dots1)}$.}

\tcb{While the cost of calibrating $\widetilde{\mathbf{M}}$ is lower bounded by the cost of estimating the cheapest column of $\mathbf{M}$, the calibration will be significantly cheaper than for $\mathbf{M}$'s most expensive column. Because full mitigation requires estimating $\sim2^{n-1}$ columns that are more sampling-expensive than the single column required for BFA and $\sim2^{n-1}$ that are cheaper, we expect the total reduction in calibration cost brought by BFA to average out to a factor of $\sim2^n$, despite the variable column cost in $\mathbf{M}$ caused by readout error biases.}

\tcb{This analysis considers a naive estimation of $\mathbf{M}$ and $\widetilde{\mathbf{M}}$, using a full dense matrix model rather than making any assumptions about correlations present. Simplified models (e.g. TPN) that require estimation of fewer parameters will incur a smaller sampling cost than the general model considered here. However, because BFA ensures symmetry in the effective response matrix thereby reducing the number of parameters needed to describe it, we expect that it will yield a reduction in calibration cost in practically all realistic cases. The bounds we use for the analysis of a generic $\mathbf{M}$ are valid for the estimation of arbitrary discrete probability distributions. Therefore they are also applicable to the sampling cost of estimating the noisy (pre-mitigation) measurement distribution for an arbitrary quantum circuit -- i.e. this requires at most $\mathcal{O}(2^n/\epsilon^2)$ shots with or without BFA.}

\begin{center}
\begin{table}
\begin{tabular}{ |m{2.4cm}||m{1.65cm}|m{1.8cm}|m{1.6cm}|  }
 \hline
 \multicolumn{4}{|c|}{Error Mitigation Scheme Summary} \\
 \hline
 Scheme& Num. measured states & Num. free parameters&Correlated errors\\
 \hhline{|=#=|=|=|}
 Full Mitigation   & $2^n$    &$2^n(2^n-1)$&Yes   \\
 \hline
 Tensor Product Noise&   $2$  & $2n$   &No\\
 \hline
 Bit-flip Averaging & $1$ & $2^n-1$& Yes\\
 \hline
 BFA+TPN    & $1$ & $n$ &  No\\
 \hline
\end{tabular}
\caption{\label{tab}A summary of the mitigation schemes considered in this paper, highlighting key characteristics.}
\end{table}
\end{center}

\subsection{Simulated Measurement of Response Matrices}\label{subsec:dev_rmat}
To obtain realistic readout error models for our simulations we measured full response matrices on $\textit{ibmq\_manhattan}$ \cite{Manhattan} for $1$ to $8$ qubit readouts (data taken on 01/12/2020). To minimize sampling error, we used $2^{16}$ shots per computational basis state (per column of $\mathbf{M}$). Taking these measured response matrices as ``exact", we used $\mathbf{M}$ to simulate the readout error process as described in Eq.~\ref{rmat}. This effectively simulates a full on-device readout process from which we benchmark various BFA strategies.  
The different schemes used for our simulations are summarised in Table~\ref{tab}.

\begin{figure}
\includegraphics[scale=0.3]{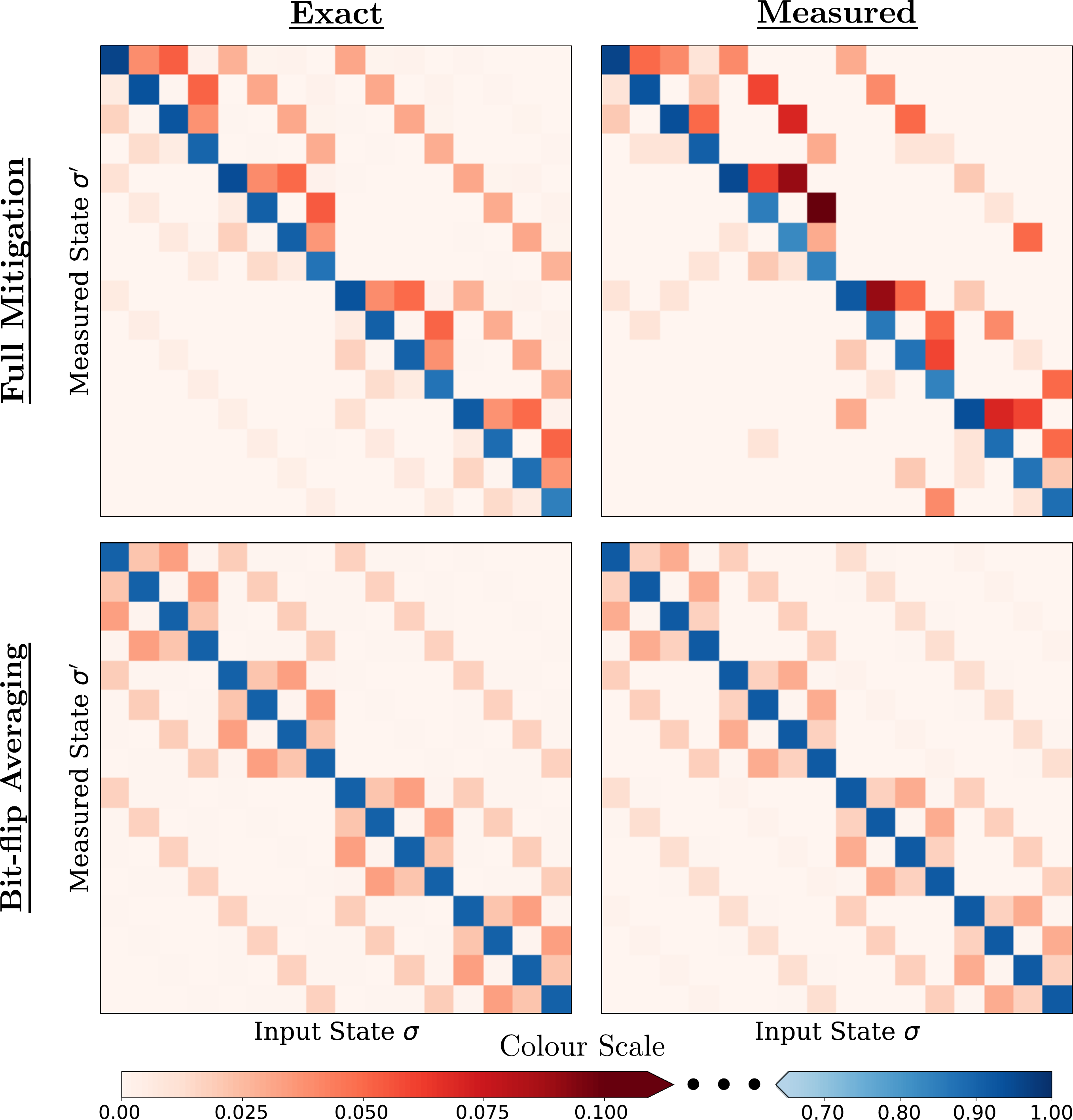}
\caption{\textbf{Example response matrix plots from 4 qubits showing the exact matrix (left), and a finite shot ($100\times 2^4$ shots) estimation (right).} For full measurement (top) the bias is manifest by lack of diagonal symmetry, and the many independent parameters increase the sampling error of its estimate. The BFA response matrix (bottom) exhibits many symmetries, and sampling error is nearly imperceptible. The exact response matrix for the error process (top left) is measured from $\textit{ibmq\_manhattan}$ using \tcb{$2^{20}$ total shots ($2^{16}$ per input state)}, and the exact BFA response matrix (bottom left) is calculated using Eq.~\ref{eq:M_BFA}.
}
\label{fig:r_mats}
\end{figure}

Figure~\ref{fig:r_mats} demonstrates the advantage of using BFA, comparing the exact and calibrated 4-qubit response matrices. For both schemes a budget of $100\times2^4$ shots were used to estimate the response matrix. Here the BFA advantage is immediately obvious: for full mitigation this budget must be divided between the $2^4$ input states that are measured while for BFA, all $100\times2^4$ shots are used to measure $\ket{0000}$ (with shot-by-shot bit-flipping). For this budget, BFA produces an accurate estimate of its target response matrix $\widetilde{\mathbf{M}}$, in contrast to full mitigation's poor estimate of $\mathbf{M}$. A more accurate response matrix allow for more effective error mitigation in the final experiment. 

\tcb{
We stress that the calibration shown in Fig. 2 assumes a full dense response matrix, rather than a TPN model. It is likely that the response matrix used for Figure~\ref{fig:r_mats} (measured on-device) can be well approximated using a TPN model allowing it to be sampled more efficiently (as only marginal distributions per qubit rather than full $2^n$-outcome distributions must be estimated). However, to showcase the advantage BFA brings over the most general approach for generic response matrices we have used the full dense model here which requires the $2^n$-outcome measurement distributions for all $2^n$ computational basis state inputs to be measured.}

Figure~\ref{fig:r_mats} shows BFA greatly reduces the amount of calibration data required to obtain a faithful description of the response matrix. We can quantify this reduction by examining the number of shots required for the estimated response matrix to converge to that of the underlying error model. To measure the ``closeness" of the estimated response matrix $\mathbf{M}_{obs}$ to the true response matrix $\mathbf{M}$ we use the average of the column-wise classical fidelities (as each column is a distinct probability distribution), which for distributions $\mathbf{p}$ and $\mathbf{q}$ is defined as $\sum_i \sqrt{p_i q_i}$. Our figure of merit $\mathcal{F_M}$ (which we will refer to as the ``response matrix fidelity") for the ``closeness" of two response matrices $\mathbf{M}$ and $\mathbf{N}$ is then given by 
\begin{equation}
\mathcal{F_M}(\mathbf{M}, \mathbf{N})=\frac{1}{2^n}\sum_{i,j} \sqrt{M_{ij}N_{ij}}.
\label{eq:MatFid}
\end{equation}

\begin{figure*}
\includegraphics[width=1.95\columnwidth]{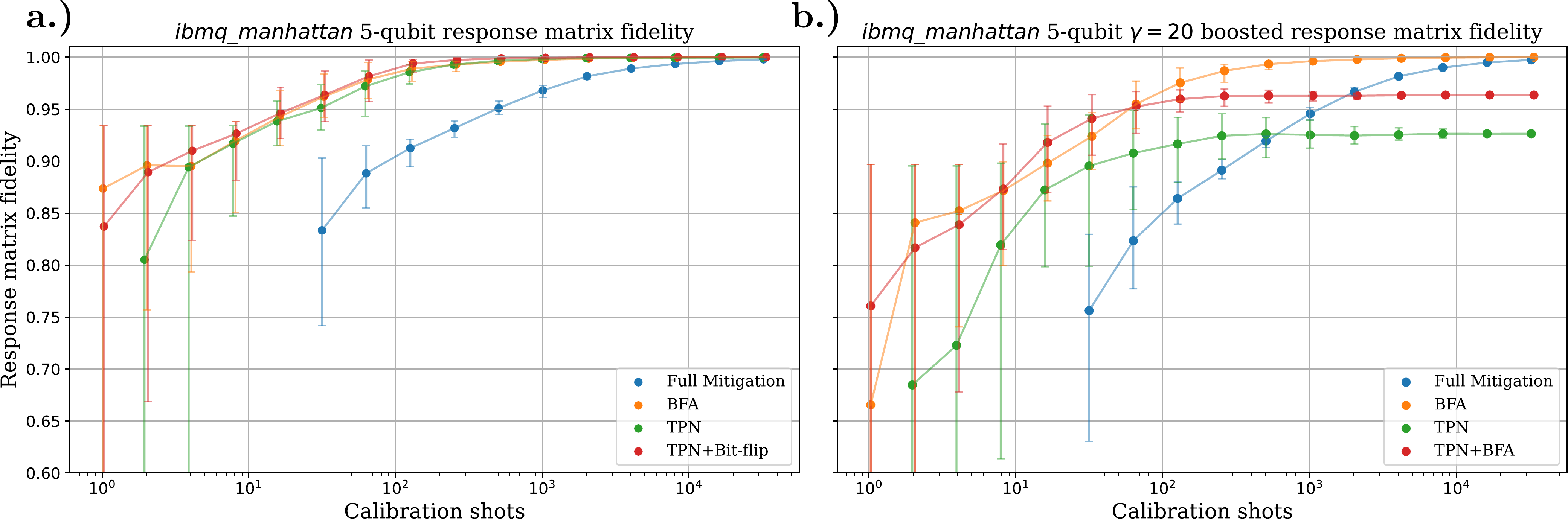}
\caption{\textbf{Scaling of the response matrix fidelities with total calibration shots.} 
\textbf{a.)} Readout errors simulated using $\mathbf{M}$ measured for $n=5$ qubits on $\textit{ibmq\_manhattan}$. 
The fidelity, Eq.~\ref{eq:MatFid} is calculated relative to the exact response matrix $\mathbf{M}$ (Full Mitigation and TPN) or $\widetilde{\mathbf{M}}$ (BFA and BFA+TPN). \textbf{b.)} As in \textbf{a.)}, using the same response matrix but with artificially boosted correlations (with boosting factor $\gamma=20$, see section~\ref{boost}). For both plots the fidelities are averaged over 50 repeats with error bars showing the middle 95\% percentile of data.}
\label{fig:manhattan_fid}
\end{figure*}

Figure~\ref{fig:manhattan_fid}a compares the fidelity between the exact and estimated response matrices for the different models considered. For a fair comparison, the same total number of shots are used to estimate each model, and the fidelities are averaged over fifty trials. The exact response matrix used here is one we directly measured for $5$-qubits on $\textit{ibmq\_manhattan}$ and so provides a realistic picture of how the schemes fare on current devices. For full mitigation and the TPN model the target is the exact response matrix used to simulate the error process while for BFA and the combination of BFA+TPN the target is the symmetrised version of the exact matrix (as in Eq.~\ref{sparse}).\par 
In comparison to the simplified schemes, full mitigation requires a far greater number of measurements to converge to a good recreation of the true response matrix. BFA very quickly converges to the maximum response matrix fidelity, requiring around two orders of magnitude fewer calibration shots than full mitigation to reach comparable fidelities. The schemes using TPN converge to $\mathcal{F_M}\rightarrow 1$, indicating the TPN assumption (i.e. independent readout errors) is justified for this particular experimentally measured response matrix. The combination of TPN and BFA yields the best fidelity with the fewest calibration shots as it is the most parameter efficient model that sufficiently captures the true readout error process. \par
As discussed in section~\ref{subsec:flip}, the same calibration measurements are required to infer the BFA and BFA+TPN response matrices. In this instance the combination of TPN+BFA was the fastest model to yield a useful description of the error process. However, in situations where significant cross-talk leads to correlated readout errors, the TPN approximation of independent errors becomes invalid. It is therefore helpful to also consider the scaling of the different schemes in a case where the TPN assumption manifestly fails.\par

\subsection{Response Matrix Measurement with Cross-talk}
To provide a toy model for cross-talk induced correlated errors we change the experimentally measured response matrix $\mathbf{M}$ to artificially amplify correlations between readout errors on adjacent qubits. 
Specifically, we boost the probability of particular syndromes to get a new response matrix $\mathbf{M}_{\gamma}$. The strength of the amplification is parameterised by a boosting factor $\gamma$ (see section~\ref{boost} for details).
Figure~\ref{fig:manhattan_fid}b shows how $\mathcal{F_M}$ scales with the number of calibration shots.The 5-qubit response matrix used here is that of Figure~\ref{fig:manhattan_fid}a but correlation boosted with $\gamma=20$. Here, $\gamma=20$ was chosen to highlight the contrast between the TPN-based schemes and those capable of dealing with correlated errors.\par

We see that in the presence of correlated errors the BFA provides a clear advantage, converging to the optimal fidelity with far fewer shots (around a factor $2^n=32$) than required for full mitigation. The two TPN-based models saturate at fidelities well below the optimal value of $1$ as they cannot account for correlated errors by construction. The fidelities obtained take longer to saturate than in Figure~\ref{fig:manhattan_fid} because the underlying readout error process with boosted correlations is more complicated than the original response matrix (which can be accurately described with TPN).\par
\begin{figure}
\includegraphics[scale=0.4]{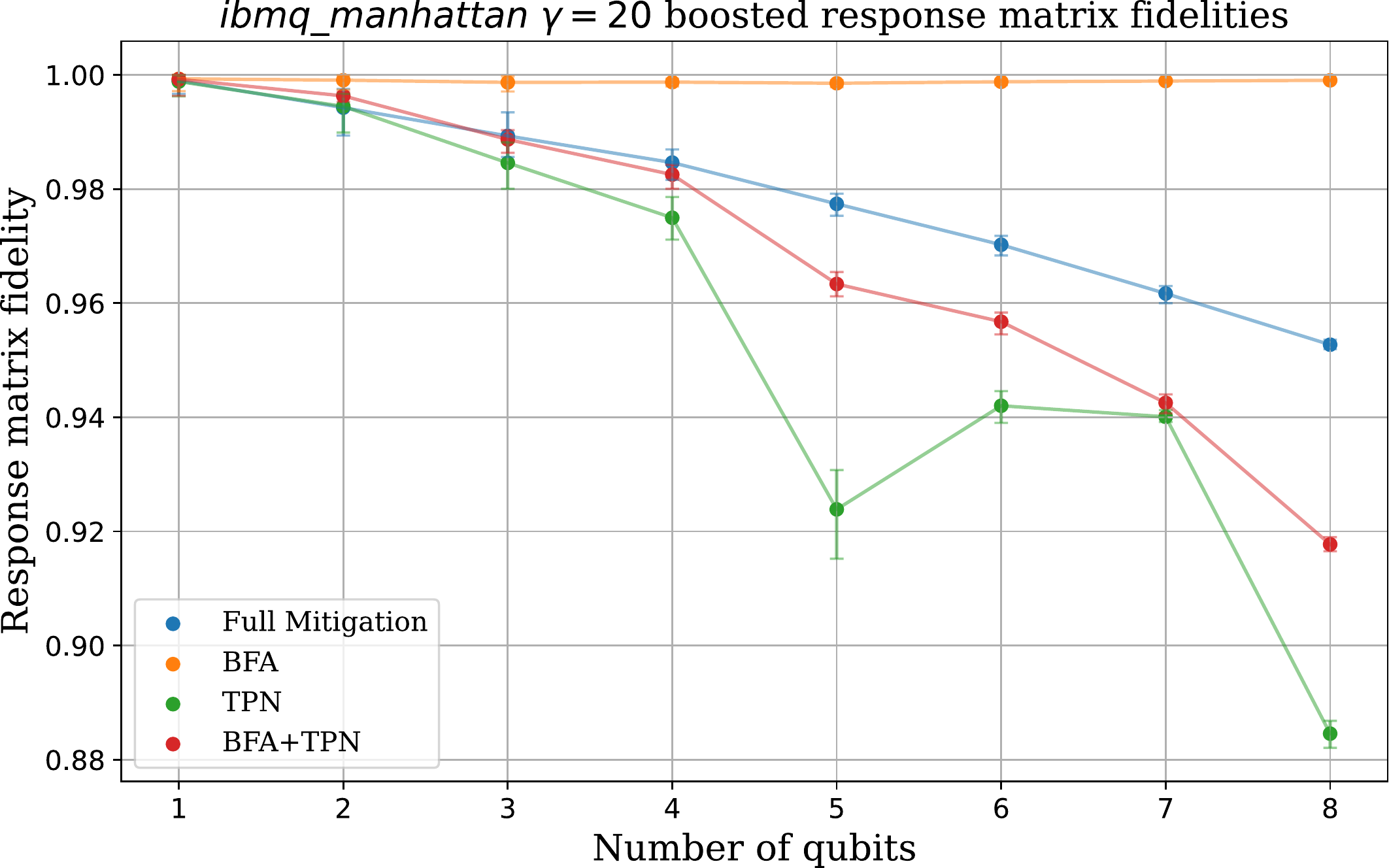}
\caption{\textbf{Demonstrating the advantage of BFA with increasing numbers of qubits $n$.} For each $n$, the response matrix was measured on $\textit{ibmq\_manhattan}$, and the correlations between neighbouring qubits boosted with $\gamma=20$. This matrix was then used to sample $2^n\times100$ shots and estimate the response matrix fidelities. The error bars (sampling error) show the middle 95\% percentile of 50 fidelity estimates.}
\label{fig:flipadv} 
\end{figure}

The advantage of BFA over full mitigation will become increasingly apparent as more qubits are measured, as demonstrated in Figure~\ref{fig:flipadv}. The plot shows the fidelity of ($\gamma=20$) boosted response matrices taken on $\textit{ibmq\_manhattan}$ estimated for different numbers of qubits. At each ($n$-qubit) point, $2^n\times 100$ simulated shots were used to estimate the response matrix.\par

Again, the full mitigation scheme has to share the $2^n\times100$ budget amongst the $2^n$ input basis states leading to increasingly severe sampling errors. 
While the TPN (and BFA+TPN) schemes have much lower sampling error, they both suffer by their inability to express correlated errors. This limit is less of a problem for BFA+TNP, as the BFA symmetrisation helps reduce the effect of correlations, and the fewer free parameters further reduce sampling errors. In particular, we note the response matrix estimated by TPN+BFA is a better approximation to the $\gamma$-boosted response matrix than is managed by TPN alone. This hints at a further advantage of the BFA scheme; averaging over different error probabilities damps biased correlations in the error model, yielding an effective error model that is closer to TPN. Finally we note that full BFA obtains by far the best fidelity as it most effectively balances sampling errors and (correlation) model expressibility.

\subsection{Simulating Graph State Fidelity Measurement}\label{subsec:experiment}
For direct comparison with previous work, we consider the example given in \cite{bravyi_2020}. We demonstrate our BFA scheme in a practical context by considering the problem of measuring the fidelity of a linear graph state of varying numbers of qubits. Again we compare combinations of full mitigation and TPN models with BFA on simulated measurements using the experimentally measured response matrices from \textit{ibmq\_manhattan}. For a linear array of $n$ qubits with initial state $\ket{+}^{\otimes n}$, a linear graph state $\ket{g_n}$ is created by applying controlled-$\hat{Z}$ gates to adjacent qubits. This graph state has a stabilizer group $\mathcal{S}^n$ generated by the set of Pauli operators 
$G_i^n = Z_{i-1} X_i Z_{i+1}$ (dropping the $Z_{-1}$ and $Z_{n}$ operators for $i=0,n$ respectively). The state fidelity can be measured averaging the expectation value of elements in $\mathcal{S}^n$. As a simplification, and to ensure we are making consistent comparisons, we only measure the generators $G^n_i$ of the stabilizer group themselves, providing an approximation to the fidelity.\par
\begin{figure*}
\includegraphics[scale=0.4]{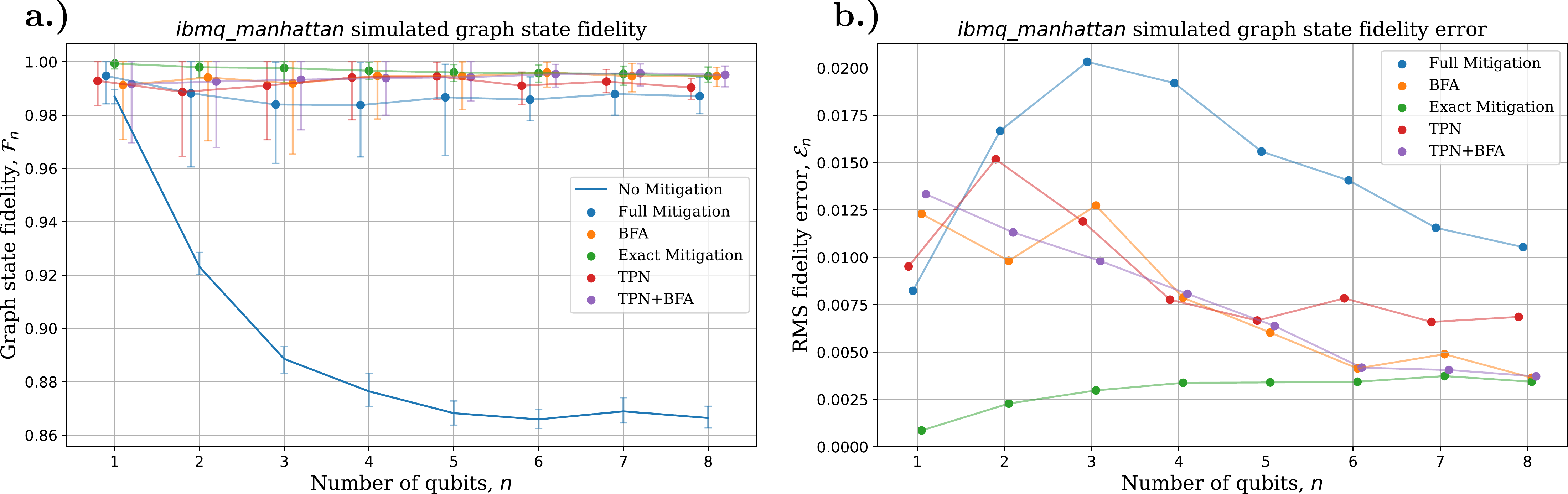}
\caption{\textbf{Demonstrating the effectiveness of readout error mitigation schemes for observable (graph state fidelity) estimation.} \textbf{a.)} Simulated estimates of graph state fidelity for varying numbers of qubits. For an $n$ qubit graph state, $100\times2^n$ calibration shots were used to estimate the response matrix and $10^5$ shots were used for the circuits to measure the fidelities. The readout errors were generated from response matrices directly measured on $\textit{ibmq\_manhattan}$. For comparison the mitigation free, and exactly mitigated (using the $\textit{ibmq\_manhattan}$ response matrices) fidelities are also shown. \textbf{b.)} Root-mean-square fidelity error $\mathcal{E}^{(model)}_{n}=(\mathbb{E}{[(\mathcal{F}^{\ (model)}_{n}-\mathbb{E}[\mathcal{F}^{\ (exact)}_{n}])^2]})^{1/2}$ with respect to mean of ``Exact Mitigation". For both plots the data is averaged over 50 repeats with error bars indicating the middle 95\% of the population.}\label{fig:graph_fid}
\end{figure*}
Figure~\ref{fig:graph_fid}a shows this approximation $\mathcal{F}_{n}=\sum_i\langle G^n_i\rangle /n$ of the fidelity found for simulations in which graph states of different sizes are prepared noiselessly and then measured with readout errors. Readout error mitigation is then applied using the four schemes. The response matrices used to simulate the readout errors were those experimentally measured on the $\textit{ibmq\_manhattan}$. For comparison, the fidelities with no mitigation applied and those for mitigation using the exact response matrix of the noise process are also shown. For the simulation of an $n$-qubit graph state a budget of $100\times2^n$ calibration shots were allowed for measurement of the model's response matrix while $10^5$ shots were used for each of the two circuits used to measure $\{G^k_i\}$.\par

The two schemes incorporating BFA perform very well, particularly with increasing system size, giving fidelities close to that achieved when mitigation is performed using the exact response matrix (i.e. perfect knowledge of the error process). However, given this calibration budget, full mitigation performs significantly worse than the other schemes. As in previous simulations, the full scheme only gets $100$ calibration shots for each of the $2^n$ input states resulting in high sampling errors.\par
It is tempting to quantify the performance of the mitigation schemes in terms how close their fidelity is to the optimal value of $1$ as we are performing simulations without gate errors. However, because we are calibrating the response matrix and calculating expectation values based on finite shots, non-correctable sampling noise remains the limiting factor. Therefore, a schemes performance should be compared against ``exact" mitigation with the response matrix used to simulate the error process. This gives the fairest estimate of the maximum improvement possible with this classical post-processing-based mitigation strategies.\par
Figure~\ref{fig:graph_fid}b shows the average root-mean-square fidelity error of the different models presented with respect to the mean ``Exact Mitigation" results, $\mathcal{E}^{(model)}_{n}=(\mathbb{E}{[(\mathcal{F}^{\ (model)}_{n}-\mathbb{E}[\mathcal{F}^{\ (exact)}_{n}])^2]})^{1/2}$, where the expectation $\mathbb{E}[\cdot]$ is taken over the fifty trials. 
Again we see that the full mitigation scheme consistently under-performs the others, suffering from far greater errors relative to ``Exact Mitigation". The TPN model performs similarly to the two bit-flipping schemes until $n> 5$, where TPN+BFA outperforms TNP alone, indicating biased correlations that BFA averages out (cf. Figures \ref{fig:flipadv} \& \ref{fig:manhattan_fid}b). For larger system sizes, both BFA-based schemes manage to replicate the ``Exact Mitigation" results with remarkable accuracy. The simplified schemes benefit from receiving exponentially many more measurement shots per calibration circuit than for full mitigation, drastically reducing sampling error. Having a more accurate response matrix means the mitigation yields results closer what would be found if using the response matrix that exactly generates the error process.\par
\section{Discussion}
We have presented bit-flip averaging (BFA), an effective scheme for readout error mitigation on near-term qubit devices. We demonstrate that BFA can augment, and consistently outperform, other measurement error mitigation strategies as it always simplifies the underlying error model. This simplification allows the response matrix to be measured using far fewer resources than would otherwise be required. Furthermore, \emph{all} BFA-augmented error models are calibrated from the same measurements, allowing these to inform the choice of model. In particular, BFA+TPN, BFA+full matrix mitigation, and all other combined schemes only require measurement of the state $\ket{0\dots 0}$.\par
BFA works by applying bit-flips to random qubits (pre-measurement) and subsequently undoing these bit-flips in the classical readout result. This greatly simplifies the observed measurement response matrix, removing all biases towards particular input states. This bias is separate from readout error correlations and so BFA does not impose any assumptions about the error process. The resulting response matrix admits a highly symmetric form. We derive a general analytic expression for its inverse and show that it can be calculated with vector-matrix multiplication. \par
We benchmark BFA using numerical simulations estimating response matrix and quantum state fidelities, and examine the role of correlated readout errors. The simulations are based on the empirical response matrices measured precisely on an IBM quantum device. Our results show that BFA can accurately estimate the response matrix with as many or fewer measurement shots than required by other schemes. Furthermore, when readout error correlations are artificially boosted, we show BFA requires orders of magnitude fewer calibration shots to find an optimal error description. Finally, we test the performance of BFA in a realistic task of measuring the (simulated) fidelity of a linear graph state in the presence of readout errors. In each case, BFA results in a more accurate fidelity estimate compared to non-bit-flipped counterparts.\par
Implementing our scheme on real devices is technically simple, but is prohibitively impractical in the current version of Qiskit. Previous works have demonstrated classical error models accurately describe on-device readout errors \cite{bravyi_2020,classical_model,classical_model2}, which encouragingly suggests that BFA will continue to surpass other mitigation strategies on physical devices. We stress that practical implementation only requires minor changes to device access, and the BFA method itself only adds (effectively free) quantum bit-flips and classical post-processing.\par
Bit-flip averaging is a useful tool for NISQ era quantum computing, allowing noisy measurements to be mitigated even in the presence of significant readout error correlations. This provides more freedom in the fabrication of solid-state quantum devices, allowing more compact qubit layouts and greater connectivity. Efficiently mitigating correlated errors is particularly important for the current generation of quantum processors where high quality devices are in high demand but short supply. Being able to perform small tasks on lower quality devices without being significantly disadvantaged by readout noise helps alleviate the throughput issues that currently limit the effectiveness of near-term quantum algorithms.
\section{Materials and Methods}
\subsection{Simulations}
Our simulations were performed using Qiskit's \textit{QasmSimulator} \cite{qiskit}. The circuits were simulated noiselessly and sampled finitely many times. These results of these measurement shots were then fed into the classical model described in Eq.~\ref{rmat} to simulate the readout error process. If a computational basis state $\ket{j}$ (expressed in binary) is measured, then the final output is sampled from the probability distribution given by the $j^\text{th}$ column of the target response matrix. 
\subsection{Derivation of inverse BFA-simplified response matrix}\label{deriv}
Starting from Eq.~\ref{sparse}, we can diagonalise $\widetilde{\mathbf{M}}$ using the Hadamard matrix:
\begin{equation}\label{diag}
\text{H}^{\otimes n}\widetilde{\mathbf{M}}\text{H}^{\otimes n}=\sum_s \tilde{p}_{(s)} Z^{(s)},
\end{equation}
where $Z^{(s)}$ is defined in the same way as $X^{(s)}$. While $\widetilde{M}$ is a classical matrix, we note its analytical form, Eq.~\ref{sparse}, lends itself to simple manipulations using the Pauli algebra. As this is now a sum over diagonal Pauli $Z$ matrices, the eigenvalues $\{\lambda_i\}$, of $\widetilde{\mathbf{M}}$ are now just the diagonal elements of this transformed matrix. To find these eigenvalues we use that the $i$th diagonal element of the Pauli operator $Z^{(s)}$ is given by
\begin{equation}
Z^{(s)}_{ii}=\prod_k(-1)^{i_ks_k}=(-1)^{i\cdot s},
\end{equation}
where $i\cdot s$ is the dot-product between binary-vector representations of the integers $i$ and $s$. This can then be rewritten in terms of the elements of the Hadamard operator $\text{H}_{ij} = (-1)^{ij}/\sqrt{2}$ giving
\begin{equation}
Z^{(s)}_{ii}=\sqrt{2^n}\text{H}^{\otimes n}_{is}
\end{equation}
The diagonal elements of $\sum_s \tilde{p}_{(s)}Z^{(s)}$, and so the eigenvalues of $\widetilde{\mathbf{M}}$, are then given by 
\begin{equation}
\begin{split}
\lambda_i&=\sum_s \tilde{p}_{(s)}Z^{(s)}_{ii}=\sum_s \tilde{p}_{(s)}(-1)^{s\cdot i}\\
&=\sqrt{2^n}\sum_s \text{H}^{\otimes n}_{is}\tilde{p}_{(s)}\\
&=\sqrt{2^n}(\text{H}^{\otimes n}\tilde{p})_i.
\end{split}
\end{equation}
Now that we have shown Eq.~\ref{eigs} and obtained the eigenvalues of $\widetilde{\mathbf{M}}$ we can find the inverse $\widetilde{\mathbf{M}}^{-1}$. We can do this by finding the projections of the diagonalised inverse matrix $\text{H}^{\otimes n}\widetilde{\mathbf{M}}^{-1}\text{H}^{\otimes n}=\Lambda^{-1}=\text{diag}(1/\lambda_0,\dots, 1/\lambda_{2^n-1})$ onto the different Pauli $Z$ matrices $\{Z^{(i)}\}$ as, after the diagonalising transformation is undone (sending $Z^{(i)}\to X^{(i)}$), these projections will give us the coefficients $\{\tilde{q}_i\}$ in Eq.~\ref{sparse_inv}. As $\text{Tr}(Z^{(i)}Z^{(j)})=2^n\delta_{ij}$, these components can be found by taking the trace of the diagonalised inverse multiplied by the different $Z^{(i)}$:
\begin{equation}
\begin{split}
\tilde{q}_{(s)}&=\frac{1}{2^n}\text{Tr}(\Lambda^{-1}Z^{(s)})=\frac{1}{2^n}\sum_{j}\frac{1}{\lambda_j}Z^{(s)}_{jj}\\
&=\frac{1}{\sqrt{2^n}}\sum_{j}\frac{\text{H}^{\otimes n}_{sj}}{\lambda_j}=\frac{1}{\sqrt{2^n}}(\text{H}^{\otimes n}\lambda^{-1})_s.
\end{split}
\end{equation}
\subsection{Boosting correlations in readout errors}\label{boost}
To amplify the errors on adjacent qubits we multiply the probability that an error occurs with syndrome $S$ by $\gamma^{n_{p}(S)}$, where $\gamma$ is a boosting factor and $n_{p}(S)$ is the number of adjacent $1$-valued bit pairs in the syndrome $S$. The response matrix is then renormalised. In our simulations we take adjacent to mean pairs of qubits that can be acted on with a two-qubit gate. The response matrices that were measured on $\textit{ibmq\_manhattan}$ were for qubits connected in a linear chain meaning that, for example, the error probability $p_{(01001|10100)}$  which has syndrome $01001\oplus 10100=11101$ would be multiplied by $\gamma^2$ as there are two pairs of qubits that have undergone readout errors. 
\begin{acknowledgments}
\textbf{Acknowledgements:}
We acknowledge the use of IBM Quantum services for this work. The views expressed are those of the authors, and do not reflect the official policy or position of IBM or the IBM Quantum team. 
\textbf{Funding:}
This work has been supported by the UK EPSRC (EP/P510257/1), the EPSRC Hub in Quantum Computing and Simulation (EP/T001062/1), the Royal Society and the Samsung GRP grant.
\textbf{Author Contributions:} AS and MSK planned the research. AS, KK, and CS discussed the initial idea which was then developed and analysed by AS. All authors discussed the results and contributed to the final manuscript. MSK supervised the project.
\textbf{Competing Interests:}
The authors are inventors on a patent application related to this work filed by Samsung Electronics Co. Ltd. (US 63/147388). The authors declare no additional competing interests.
\textbf{Data and materials
availability:}
All data needed to evaluate the conclusions in the paper are present in the
paper and/or the Supplementary Materials. \tcb{The raw data used to generate the plots (Figures 3-5) and the raw response matrix calibration data from $\textit{ibmq\_manhattan}$ can be found at https://doi.org/10.5281/zenodo.5267397}. 
\end{acknowledgments}
\bibliography{refs}




\setcounter{equation}{0}
\setcounter{section}{0}
\def\theequation{S\arabic{equation}}
\setcounter{table}{0}
\renewcommand{\thetable}{\textbf{S}\Roman{table}}
\renewcommand{\thesection}{\textbf{S}\Roman{section}}
\makeatletter
\newcommand{\globalcolor}[1]{%
  \color{#1}\global\let\default@color\current@color
}
\makeatother


\newpage
\begin{center}
\large{\textbf{Supplementary Text}}
\end{center}
\tcb{
\section{Example of mitigation with sparse representation}\label{appendix}
Here we give an illustrative example of how BFA can simplify readout error mitigation. Suppose we have a four-qubit system in which errors on two qubits are completely correlated (i.e. when an error occurs on one qubit it always occurs on the other). An example of the underlying response matrix for this system could be:
\begin{equation}\label{appendix:M}
    \mathbf{M}=\underbrace{\begin{pmatrix}
    0.98&0.08\\
    0.02&0.92
    \end{pmatrix}}_{\text{qubit 3}}
    \otimes
    \underbrace{\begin{pmatrix}
    0.96&0&0&0.16\\
    0&0.94&0.1&0\\
    0&0.06&0.9&0\\
    0.04&0&0&0.84
    \end{pmatrix}}_{\text{qubit 2 and 1}}
    \otimes
    \underbrace{\begin{pmatrix}
    0.97&0.11\\
    0.03&0.89
    \end{pmatrix}}_{\text{qubit 0}}
\end{equation}
Under BFA the symmetrised form of this matrix (using the sparse representation in Eq. 9) is:
\begin{equation}\label{appendix:Mt}\begin{split}
    \widetilde{\mathbf{M}}=&\underbrace{(0.95\mathbb{1}+0.05X)}_{\text{qubit 3}}\otimes\underbrace{(0.91\mathbb{1}\mathbb{1}+0.09XX)}_{\text{qubit 2 and 1}}\otimes\underbrace{(0.93\mathbb{1}+0.07X)}_{\text{qubit 0}}\\
    =&\ 0.804\mathbb{1}\mathbb{1}\mathbb{1}\mathbb{1}+0.06\mathbb{1}\mathbb{1}\mathbb{1}X+0.8\mathbb{1}XX\mathbb{1}\\
    &+0.006\mathbb{1}XXX+0.042X\mathbb{1}\mathbb{1}\mathbb{1}+0.003X\mathbb{1}\mathbb{1}X\\
    &+0.004XXX\mathbb{1}+0.0003XXXX,
\end{split}
\end{equation}
where, as discussed in section II A, the coefficients in front of the operators in this expansion are the probabilities that errors of a the corresponding syndrome occur (under bit-flipping). Suppose we now take a set of calibration measurements, inputting the state $\ket{0000}$, applying random bit-flips before a measurement and correcting the result to undo the bitflip. An example set of calibration results for this (with 10,000 repeats), showing the observed frequencies/probabilities for the different outcomes and their true values (from the underlying response matrix), is shown in Table~\ref{calib_eg}.}
\tcb{
\begin{table}[h!]
\begin{center}
\begin{tabular}{ |m{2cm}||m{1.65cm}|m{1.65cm}|m{1.6cm}|  }
 \hline
 \multicolumn{4}{|c|}{BFA Simulated Calibration Results} \\
 \hline
 Outcome& Obs. freq. & Est. prob.&True prob.\\
 \hhline{|=#=|=|=|}
 0000   & 8091 & 0.8091 & 0.8040   \\
 \hline
 0001   & 595 & 0.0595 & 0.0605   \\
 \hline
 0110   & 784 & 0.0784 & 0.0795   \\
 \hline
 0111   & 61 & 0.0061 & 0.0060   \\
 \hline
 1000   & 433 & 0.0433 & 0.0423   \\
 \hline
 1001   & 22 & 0.0022 & 0.0032   \\
 \hline
 1110   & 46 & 0.0046 & 0.0042   \\
 \hline
 1111   & 4 & 0.0004 & 0.0003  \\
 \hline
\end{tabular}
\end{center}
\caption{\textbf{Example simulated calibration results under BFA for the response matrix given in Eq.~\ref{appendix:M}.} Shots (10,000) are sampled randomly from the columns in the underlying $\mathbf{M}$ (equivalent to bit-flipping an input state $\ket{0\dots0}$) and a classical correction is applied.}\label{calib_eg}
\end{table}}
\tcb{
This immediately provides us with an estimate $\widetilde{\mathbf{M}}^*$ for the BFA-simplified response matrix $\widetilde{\mathbf{M}}$ as
\begin{equation}\label{app:est}\begin{split}
    \widetilde{\mathbf{M}}^*=&\ 0.8091\mathbb{1}\mathbb{1}\mathbb{1}\mathbb{1}+0.0595\mathbb{1}\mathbb{1}\mathbb{1}X+0.784\mathbb{1}XX\mathbb{1}\\
    &+0.0061\mathbb{1}XXX+0.0433X\mathbb{1}\mathbb{1}\mathbb{1}+0.0022X\mathbb{1}\mathbb{1}X\\
    &+0.0046XXX\mathbb{1}+0.0004XXXX.
\end{split}
\end{equation}
This has an infidelity ($1-\mathcal{F}_\mathcal{M}$) with the true response matrix of $\approx10^{-4}$. However, this estimate does not necessarily exactly admit the decomposition in Eq.~\ref{appendix:Mt} (qubits 0 and 3 having independent errors and qubits 1 and 2 correlated errors). If we had prior knowledge that such a decomposition was likely to be correct then we could instead assume this decomposition and infer the probabilities by looking at the marginal results, shown in Table~\ref{marginals}.
\begin{table}[h!]
\begin{center}
\begin{tabular}{ |m{2cm}||m{1.65cm}|m{1.65cm}|m{1.6cm}|  }
 \hline
 \multicolumn{4}{|c|}{Qubit 0 BFA Calibration Marginals} \\
 \hline
 Outcome& Obs. freq. & Est. prob.&True prob.\\
 \hhline{|=#=|=|=|}
 0   & 9318 & 0.9318 & 0.93   \\
 \hline
 1   & 682 & 0.0682 & 0.07   \\
 \hline
 \multicolumn{4}{|c|}{Qubits 2 and 1 BFA Calibration Marginals} \\
 \hline
 Outcome& Obs. freq. & Est. prob.&True prob.\\
 \hhline{|=#=|=|=|}
 00   & 9141 & 0.9141 & 0.91   \\
 \hline
 01   & 0 & 0.0 & 0.0   \\
 \hline
 10   & 0 & 0.0 & 0.0   \\
 \hline
 11   & 859 & 0.0859 & 0.09   \\
 \hline
 \multicolumn{4}{|c|}{Qubit 3 BFA Calibration Marginals} \\
 \hline
 Outcome& Obs. freq. & Est. prob.&True prob.\\
 \hhline{|=#=|=|=|}
 0   & 9495 & 0.9495 & 0.95   \\
 \hline
 1   & 505 & 0.0505 & 0.05   \\
 \hline
\end{tabular}
\caption{\textbf{Marginal simulated calibration results under BFA for the response matrix in Eq.~\ref{appendix:M}.} These results are those shown in Table~\ref{calib_eg} marginalised on the individual qubits 0 and 3 and the pair of qubits 1 \& 2 that undergo correlated errors.}\label{marginals}
\end{center}
\end{table}}
\tcb{
By making this assumption about the structure of the readout errors we reach a slightly different estimate for the response matrix:
\begin{equation}\begin{split}
    \widetilde{\mathbf{M}}^*=\underbrace{(0.945\mathbb{1}+0.05X)}_{\text{qubit 3}}&\otimes\underbrace{(0.9141\mathbb{1}\mathbb{1}+0.0859XX)}_{\text{qubit 2 and 1}}\\
    &\otimes\underbrace{(0.9318\mathbb{1}+0.0682X)}_{\text{qubit 0}},
\end{split}
\end{equation}
which has a lower infidelity of $\approx3\times10^{-5}$ and so will result in more accurate mitigation. A tensor product structure such as this greatly reduces the computational cost of finding the inverse response matrix (which is the tensor product of the inverse matrices for qubit 3, qubits 2 and 1, and qubit 0 respectively). In this case keeping the more general structure in Eq.~\ref{app:est} also allows the inverse to be found more easily as the Pauli matrices in this expansion form a closed group under matrix multiplication. This means that $(\widetilde{\mathbf{M}}^*)^{-1}$ must contain the same set of $X$ operators as $\widetilde{\mathbf{M}}^*$ ($\mathbbm{1}\mathbbm{1}\mathbbm{1}\mathbbm{1}, \mathbbm{1}\mathbbm{1}\mathbbm{1}X,$ etc.) meaning that fewer coefficients need to be calculated than for a generic $\widetilde{\mathbf{M}}^{-1}$.}

\tcb{
The constrained optimization approach to mitigation is also made easier as the response matrix is relatively sparse. For measurements of some states this is made easier still as we can use a reduced set of possible outcomes. As an example we consider the noisy readout of a noiseless preparation of the 4-qubit GHZ state $(\ket{0000}+\ket{1111})/\sqrt{2}$. If the underlying response matrix that in Eq.~\ref{appendix:M}, a set of possible outcome measurements (probabilities from 10,000 shots) while using BFA is given in Table~\ref{ghz_res}.
\begin{table}[h!]
\begin{center}
\begin{tabular}{ |m{2cm}||m{1.6cm}|m{1.8cm}|m{1.6cm}|  }
 \hline
 \multicolumn{4}{|c|}{BFA 4-qubit GHZ State Simulated Results} \\
 \hline
 Outcome& Observed prob. & Expected noisy prob.&Noiseless prob.\\
 \hhline{|=#=|=|=|}
 0000   & 0.4048 & 0.40215 & 0.5   \\
 \hline
 0001   & 0.0330 & 0.03235 & 0.0   \\
 \hline
 0110   & 0.0407 & 0.04135 & 0.0   \\
 \hline
 0111   & 0.0243 & 0.02415 & 0.0   \\
 \hline
 1000   & 0.0235 & 0.02415 & 0.0   \\
 \hline
 1001   & 0.0414 & 0.04135 & 0.0   \\
 \hline
 1110   & 0.0316 & 0.03235 & 0.0   \\
 \hline
 1111   & 0.4007 & 0.40215 & 0.5  \\
 \hline
\end{tabular}
\end{center}
\caption{\textbf{Simulated GHZ state measurement results under BFA.} These probabilities are calculated from 10,000 shots. BFA is simulated by applying shot-wise random bit-flips to noiseless simulated results, sampling from the corresponding column of the underlying response matrix (in this case that in Eq.~\ref{appendix:M}), and applying a classical correction (the bit-flip again) to the result.}\label{ghz_res}
\end{table}}

\tcb{
The expected probabilities are given by Eq. 6 for this GHZ state and the noiseless probabilities are in the absence of any readout errors. We see that for this GHZ state and readout error profile we only have 8 outcomes in the observed results (as the errors map the two components of the GHZ state to the same set of outcomes). Errors with syndromes in the set $\{0000,0001,0110,0111,1000,1001,1110,1111\}$ (those observed in calibration) map the outcomes in the observed data for the GHZ state to the same set. This means that we can perform the constrained minimization mitigation while considering only this reduced set of 8 outcomes as opposed to the usual 16.}

\end{document}